# Bayesian inference for precise and uncertainty-quantified single-shot widefield interferometric geometrical nanometrology

[1,4],‡Damian M. Suski, [2,5],‡Maria Cywińska, [2]Julianna Winnik, [2]Michał Józwik, [2]Piotr Zdańkowski, [3]Azeem Ahmad, [3]Balpreet S. Ahluwalia, [2,*]Maciej Trusiak

[1] Warsaw University of Technology, Institute of Automatic Control and Robotics,

8 Sw. A. Boboli St., 02-525, Warsaw, Poland

[2] Warsaw University of Technology, Institute of Micromechanics and Photonics,

8 Sw. A. Boboli St., 02-525, Warsaw, Poland

[3] UiT The Arctic University of Norway, Department of Physics and Technology,

35 Klokkargårdsbakken St., 9019, Tromsø, Norway

[4] damian.suski@pw.edu.pl

[5] maria.cywinska@pw.edu.pl

* maciej.trusiak@pw.edu.pl

‡authors contributed equally

**ABSTRACT:** Advanced geometrical nanometrology is critical for process control in semiconductor manufacturing, supporting applications in, e.g., photonic integrated circuits, nanoelectronics, and emerging quantum and optoelectronic technologies. Widefield interferometric approach provide a cost-effective, non-destructive solution for characterizing semiconductor optical waveguides, which are fundamental to nanophotonic devices. This work presents a Bayesian inference framework, implemented using Dynamic Nested Sampling, for estimating geometric parameters - such as width and height - of a semiconductor optical waveguide from a single widefield interferogram. The proposed framework reduces the need of leveraging near field scanning microscopy methods for measurements. The notable advantage is that Bayesian statistics not only provide the estimated parameter values but also quantify the uncertainty of the inference results and the fitness of the used model. The proposed full-field, single-shot interferometric approach, supported by Bayesian-based data analysis, achieves high accuracy and sensitivity - down to successful measurement of 8 nm rib waveguide - while remaining resilient to noise. Thus, the demonstrated methodology provides a cost-effective, robust, and scalable tool for semiconductor fabrication monitoring and process verification, as confirmed by both numerical simulations and experimental validation on optical waveguides. This method contributes to high-precision nanometrology by integrating advanced statistical modeling and inference techniques.

KEYWORDS: optical metrology, nanometrology, interferometric microscopy, single-shot interferometry, Bayesian inference, model-based parameter estimation, semiconductor optical waveguides.

**INTRODUCTION**

The continuous miniaturization of semiconductor devices demands ever-higher precision in fabrication [1], making advanced metrology a cornerstone of next-generation silicon photonics (photonic integrated circuits, PICs) [2, 3], nanoelectronics [4], and quantum technologies [5,6]. Semiconductor optical waveguides, in particular, are crucial components in, e.g., high-speed optical communication [7], PICs enabled optical nanoscopy [8,9], neuromorphic computing [10], and holographic displays [11], including metasurface waveguides [12], etc. Even nanometric deviations in the waveguide geometry can lead to significant optical losses or functional impairments. As manufacturing tolerances shrink, non-contact and scalable metrological solutions become essential to ensure correct performance. However, existing characterization techniques remain limited in their ability to provide widefield, accurate, and robust measurements across large areas in a production environment.

Traditional near-field methods [1], such as atomic force microscopy and scanning electron microscopy, provide high spatial resolution but are inherently slow, requiring point-by-point scanning that limits throughput. Furthermore, these methods are often destructive, particularly for fragile photonic structures, and require specialized sample preparation, making them impractical for high-throughput metrological verification of nanofabrication in semiconductor foundries. Optical techniques in general [13], and quantitative phase imaging in particular [14-17], offer a fast, non-invasive alternative capable of characterizing nanoscale geometries across large fields of view. Conventional white light interferometric techniques are generally well-established and very capable, yet they need scanning and are limited in spatial resolution due to physically low-pass filtering microscope objectives [14]. Recently, non-interferometric phase measurement via high resolution reflective Fourier ptychographic microscopy method has been proposed [15], which is, however, susceptible to numerical reconstruction errors, and generally do not fully transfer low spatial frequencies of phase distribution.

To address these crucial challenges, we introduce a Bayesian inference framework implemented with Dynamic Nested Sampling (DNS) [18] to enable precise single-shot widefield interferometric nanometrology of semiconductor optical waveguides. Our approach extracts waveguide dimensions such as height and width, directly from a single interferogram without the need for explicit phase reconstruction. Unlike conventional phase-reconstruction-based metrology, where noise can uncontrollably transfer from the intensity to the phase domain, Bayesian inference naturally models complex parameter distributions, ensuring accurate parameter estimation with rigorous uncertainty quantification and improved robustness against noise and systematic errors. This advantage is particularly crucial for semiconductor nanofabrication control, where quantifying uncertainty in measurements directly impacts yield optimization and defect detection. In particular, the proposed framework is superior to the conventional phase-shifting interferometry approaches, to accurately measure extremely small waveguide heights, that induce only a subpixel shift in the fringes, and to perform reliably for interferograms with a very low SNR or very low number of carrier fringes (up to the uniform field mode).

Bayesian methods are increasingly utilized in optical measurements, thereby enhancing accuracy and facilitating the quantification of uncertainty. Examples of leveraging the Bayesian framework for estimating an object's geometrical parameters include interferometric measurements of lens geometry [19] and scatterometric measurements of nanostructures [20]. Bayesian methods combined with deep-learning models have been applied to fringe-pattern based phase reconstruction [21], phase imaging using non-interferometric techniques [22], and optical coherence elastography [23]. Furthermore, Bayesian frameworks have played a role in optical coherence tomography for increased depth sensitivity [24], speckle noise reduction [25], and enhanced velocimetry [26]. Other applications of Bayesian inference include estimating the distribution of particle sizes in aerosols [27] and analysis of gravitational waves [28].

In the mentioned works, the measurement and computation schemes still rely on explicit phase retrieval [19,20], scanning/multi-shot approaches [19], or deep-learning-based inference [21-23]. To our best knowledge, we propose for the first time the Bayesian inference framework, which measures the geometric parameters of the nanostructures from a single widefield interferogram. To ensure high accuracy of numerical calculations, we use DNS instead of more popular but less accurate Markov Chain Monte Carlo (MCMC) algorithms [29]. MCMC generates posterior samples through random walks, which makes it powerful but often slow and unsuitable for estimating model evidence or handling complex, multimodal distributions. Nested sampling approaches instead estimate both the posterior and model evidence using a set of evolving live points, while dynamic nested sampling further adapts by dynamic allocation of the number of live points during inference to improve efficiency and accuracy in challenging problems. It is worth emphasizing that DNS has not been used for optical measurements yet, and in this contribution, we harvest for the first time DNS's rich potential showcased in astronomical scale (black hole imaging [30]) successfully transferring it into a nanoscale.

In the Results and Discussion section, we present the physical model of the widefield nano-measurement and discuss the Bayesian inference framework. Next, we show the results of simulation experiments, which aim at assessing the validity and efficiency of the proposed approach for different measurement conditions. Afterward, we present the successful applications of our framework to the measurement of two samples – an etched calibrated nanostructure and a suspended waveguide. Finally, we discuss the obtained results and formulate the conclusions.

**RESULTS**

Our goal is to measure the geometrical parameters of the object (e.g. optical waveguide), on the basis of a single fringe pattern. The measurement data $D$ is the fringe intensity $I_D(x, y)$ registered at the image pixels with positions $(x, y)$. The measurement process is realized by estimating the parameters $\Theta$ of a fringe pattern model $M$, such that the modelled intensity pattern $I_M(x, y; \Theta)$ fits $I_D(x, y)$. The better the agreement between $I_D(x, y)$ and $I_M(x, y; \Theta)$, the stronger the confidence that the model parameters $\Theta$ are accurate. The Bayesian inference framework describes that observation quantitatively. In the next subsection, we will describe the physical model of the fringe pattern. Afterward, we will describe the Bayesian inference framework in the context of the parameter estimation task and discuss briefly DNS algorithm.

PHYSICAL MODEL OF THE WIDEFIELD NANO-MEASUREMENT. To model the waveguide geometry, we assume a simple step model presented in Figure 1(a, b). The parameters of the step model are $p_x, p_y$ – positions along $x$ and $y$ axis, $\beta$ – orientation, $w$ – width, $h$ – height, $l$ – base level. The formula describing the height map of the sample can be expressed as:

$$H(x, y; \Theta_s) = l + h \cdot S(x, y; \Theta_s), \tag{1}$$

where $\Theta_s = [p_x, p_y, \beta, w]$ is a vector of the step parameters, and $S(x, y; \Theta_s)$ is a logistic step function ($c$ – smoothing parameter):

$$S(x, y; \Theta_s) = \frac{1}{1+\exp\left(c \cdot \left(d^2(x,y;\Theta_s) - \left(\frac{w}{2}\right)^2\right)\right)}. \tag{2}$$

The step function depends on the signed distance of the point $(x, y)$ from the center line of the step (dash-dot line in Fig 1(a):

$$d(x, y; \Theta_s) = -\cos\beta \cdot (x - p_x) + \sin\beta \cdot (y - p_y). \tag{3}$$

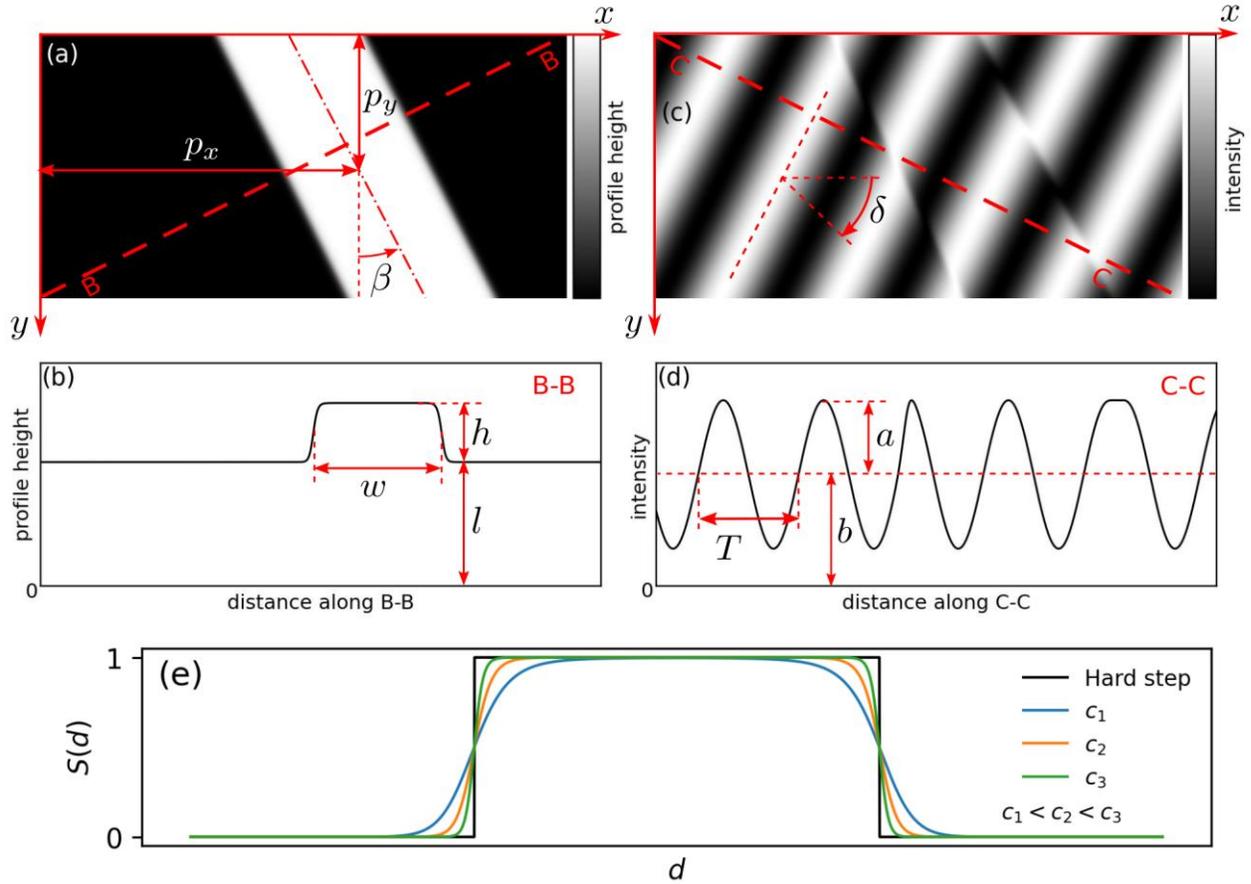

Figure 1. The description of the step and fringe model parameters. (a) Height map of the sample: $p_x, p_y$ – step positions along $x$ and $y$ axis, $\beta$ – step orientation. Center line of the step is marked with a dash-dot line. (b) Diagonal cross section of the height map: $w$ – step width, $h$ – step height, $l$ – base level. (c) Fringe intensity pattern: $\delta$ – fringe orientation. (d) Diagonal cross section of the intensity pattern: $T$ – fringe period, $a$ – fringe profile amplitude, $b$ – background intensity. (e) Comparison of the hard step and logistic step functions of distance $S(d)$ (Eq. 2) for different values of smoothing parameter $c$.

The logistic step function defined in Eq. 2 is a smoothed version of a hard step function, which returns a value of 1 for points belonging to the step and zero otherwise. The strength of the smoothing depends on the value of $c$ (see Figure 1(e)).

Applying the smoothed step function mitigates the ill-posedness of the hard step model, which emerges when the $(x, y)$-continuous model is discretized into a set of pixels. However, the smoothing parameter should be adjusted carefully. Our observation is that the smoothing parameter should be chosen such that the width of the transition region around the step edge is 2 to 10 pixels.

It is noteworthy that in the definition of the distance function (Eq. 3), the choice of the step position $(p_x, p_y)$ is ambiguous. One can replace $(p_x, p_y)$ with any point $(x_c, y_c)$ lying on the center line of the step and the distance function will give exactly the same values (see Supporting Information S1). The ambiguity in choosing the step position $(p_x, p_y)$ makes the parameter estimation task an ill-posed problem. To resolve that issue, during the computations, we set one of the positions ($p_x$ or $p_y$) to a fixed value and estimate only the other one.

We model the fringe pattern as a sinusoidal modulation with the parameters described in Figure 1(c, d): $T$ – period, $\delta$ – orientation, $a$ – fringe profile amplitude, $b$ – background intensity. We assume the following model of the fringe intensity pattern:

$$I_M(x, y; \Theta) = b + a \cdot \cos\left(\frac{2\pi}{T}(x \cdot \cos\delta + y \cdot \sin\delta) + \varphi_l + \varphi_h \cdot S(x, y; \Theta_s)\right), \quad (4)$$

where $\varphi_h$ is a phase shift induced by the step height, $\varphi_l$ is the initial phase shift of the fringe pattern and $\Theta = [p_x, p_y, \beta, w, \varphi_h, \varphi_l, T, \delta, a, b]$ is the vector of fringe model parameters. The phase shift $\varphi_h$ is directly related to the step height $h$. The exact form of that relation depends on the measurement system configuration, e.g. it is different for transmissive and reflective measurement setups.

BAYESIAN INFERENCE. The Bayesian inference is based on the famous Bayesian rule [18]:

$$P(\Theta|D, M) = \frac{P(D|\Theta,M) \cdot P(\Theta|M)}{P(D|M)} = \frac{\mathcal{L}(\Theta) \cdot \Pi(\Theta)}{\mathcal{Z}}, \quad (5)$$

where $\Theta$ are the parameters of the model $M$ and $D$ is data. Let us describe the subsequent terms in the Bayesian formula, namely $P(\Theta|D, M)$ – posterior probability, $\mathcal{L}(\Theta) = P(D|\Theta, M)$ – likelihood, $\Pi(\Theta) = P(\Theta|M)$ – prior probability and $\mathcal{Z} = P(D|M)$ – evidence [18].

We aim at calculating $P(\Theta|D, M)$ – the multidimensional posterior probability distribution of the model parameters for the given data and the assumed model. From the posterior distribution we can easily obtain the estimated parameters of the model, for example by calculating the distribution mean or median. Also, the posterior probability distribution allows for a direct calculation of the confidence bounds for the estimated parameters, e.g., as the lower and upper quantiles of the distribution. Estimated parameters together with their confidence bounds form the measurement result.

The likelihood term $\mathcal{L}(\Theta) = P(D|\Theta, M)$ expresses the probability that the given data can be an output of the model $M$ with parameters $\Theta$. Assuming that the regression residual at each pixel is normally distributed with a zero mean and the standard deviation $\sigma$:

$$e_I(x, y; \Theta) = I_M(x, y; \Theta) - I_D(x, y) \sim \mathcal{N}(0, \sigma^2) \quad (6)$$

and the residuals in different pixels are independent, the likelihood takes the form (see Supporting Information S2):

$$\mathcal{L}(\Theta) = \exp\left(-\frac{N}{2} \cdot (\log(2\pi\sigma^2) + 1)\right), \quad (7)$$

where $N = s_x \cdot s_y$ is a number of pixels in the image of size $s_x \times s_y$ and $\sigma^2$ is equal to the mean squared residual (MSR):

$$\sigma^2 = \frac{1}{N}\sum_{x,y} e_I^2(x, y; \Theta). \quad (8)$$

From the monotonicity of the exponential function in Eq. 7, it follows that the likelihood reaches maximum when the regression MSR is minimized.

The term $\Pi(\Theta) = P(\Theta|M)$ in the Bayesian formula is called prior probability, and it represents a priori knowledge about the possible values of model parameters. That knowledge can come, e.g., from previous measurements or the design assumptions for the measured object. In the former case, the prior can be modelled for example as a Gaussian distribution with the mean value and standard deviation resulting from previous measurements. In the latter case, the prior can be modelled as a uniform distribution over the interval $[\Theta_{min}, \Theta_{max}]$ representing the design tolerances for the object's geometry parameters. The uniform distribution can also be used if there are no specific sources of prior knowledge about the parameters' values. In such a case, the prior interval represents a crude estimate of the possible parameters' values. In any case, the prior distribution should be carefully adjusted, as taking excessively wide prior space results in a significant increase of the computation time [18].

The evidence $\mathcal{Z} = P(D|M)$ is defined as:

$$\mathcal{Z} = P(D|M) = \int_{\Omega_\Theta} \mathcal{L}(\Theta) \cdot \Pi(\Theta) d\Theta, \tag{9}$$

where the integral is taken over the entire domain $\Omega_\Theta$ of the model parameters. The product of likelihood and prior probability $\mathcal{L}(\Theta) \cdot \Pi(\Theta)$, which appears inside the integral is called posterior mass. The evidence can be understood as the "model probability," i.e., the overall probability that data $D$ can be an output of the model $M$, considering only the structure of the model but not the specific parameters values [18]. The evidence can be used to compare the quality of different models – the better suited models achieve higher evidence values.

The accurate computation of the evidence term is a central idea of the Nested Sampling algorithms [18]. To accurately calculate the evidence integral (Eq. 9), the samples drawn by the algorithm must be chosen carefully, such that the distribution of the posterior mass $\mathcal{L}(\Theta) \cdot \Pi(\Theta)$ over the entire parameters domain $\Omega_\Theta$ is well represented. To achieve that goal, at each iteration the algorithm estimates the region in the prior domain from which the new samples must be drawn to systematically decrease the evidence estimation error and that allows to formulate the precise convergence criteria for the algorithm. As a byproduct of the accurate estimation of the posterior mass distribution over $\Omega_\Theta$, the posterior probability (proportional to posterior mass) is also calculated accurately.

The Nested Sampling differs from the MCMC approach, where the samples are always drawn from the entire prior domain and post-selected based on their posterior mass. The MCMC algorithms are generally faster than Nested sampling algorithms due to a simpler drawing scheme, but they lack well-motivated convergence criteria [18]. The DNS algorithm further improves the computations of the posterior, by adaptively (dynamically) allocating more samples to the regions that contribute most to the posterior. For computations we use an open-source Python implementation of DNS provided with *dynesty* library [18].

Figure 2 summarizes the flowchart of the utilized numerical algorithm. Based on the measurement data, the assumed model, and the prior distribution of model parameters, DNS accurately calculates the evidence and the posterior distribution. The posterior distribution enables us calculate the estimated values of model parameters, along with their confidence bounds.

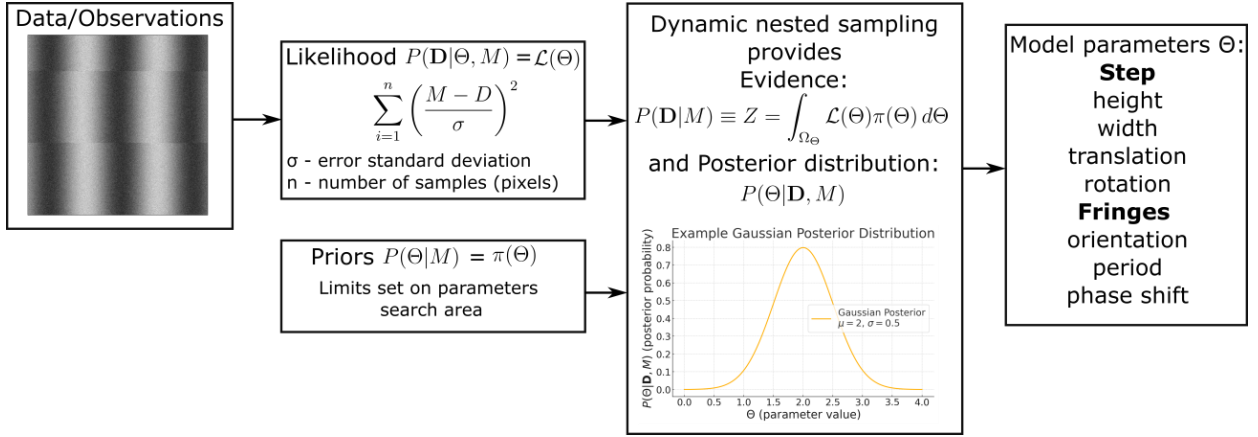

*Figure 2. Flowchart of the Bayesian inference computational framework. The inputs are the measured fringe pattern and the prior distributions (uniform distribution on a predefined prior interval in our case). The likelihood function definition is related to the fringe model regression MSR. DNS draws samples from the prior space to accurately calculate evidence and multidimensional posterior distribution of the model parameters. From the posterior distribution the estimated values of model parameters as well as their confidence bounds can be calculated, forming together the measurement result.*

SIMULATION RESULTS. The test image size is 100×100 pixels (px). We assumed a roughly vertical, centered ($p_x = p_y = 50$ px) step object of width $w = 40$ px with a small vertical tilt ($\beta = 2°$). Fringes were modelled as a horizontal sinusoidal intensity pattern (also slightly tilted $\delta = 92°$) with amplitude $a = 30$, background intensity $b = 130$, and phase shift $\varphi_l = 90°$. Additionally, a Gaussian noise signal $\mathcal{N}(0, \sigma^2)$ was added to the simulated fringe pattern. We assumed the reflective measurement configuration, thus the step height is related to the phase shift via:

$$h = \frac{\lambda \cdot \varphi_h}{4\pi}, \qquad (10)$$

where $\lambda = 635$ nm is the wavelength of the illumination light.

In Table 1, the prior intervals of the fringe model parameters are presented. The intervals were chosen on the basis of the rapid visual inspection of the simulated fringe patterns. To avoid the ambiguity problem (see Supporting Information S1) for the step position $(p_x, p_y)$, we set $p_y$ to a constant value 50 px. The simulated values of the phase shift $\varphi_{h,sim}$ and the fringe period $T_{sim}$ varied across the experiments, so the prior bounds for those parameters were also shifted accordingly.

*Table 1. Prior intervals for model parameters.*

| Parameter | $\varphi_h$[rad] | $w$[px] | $p_x$[px] | $p_y$[px] | $\beta$[°] |
|---|---|---|---|---|---|
| **Prior interval** | $\varphi_{h,sim} \pm \frac{\pi}{4}$ | $40 \pm 5$ | $50 \pm 5$ | $const = 50$ | $2 \pm 5$ |
| **Parameter** | $\delta$[°] | $\varphi_l$[rad] | $T$[px] | $a$[a.u.] | $b$[a.u.] |
| **Prior interval** | $92 \pm 5$ | $[0, \pi]$ | $T_{sim} \pm 5$ | $30 \pm 5$ | $130 \pm 5$ |

First, we have evaluated the DNS performance for different step heights. We assumed the period $T = 32$ px, and the Gaussian noise level $\sigma = 0.1$. The obtained results are presented in Figure 3.

The output of DNS is the estimated multidimensional posterior probability distribution of the parameters vector over the prior space. In Figure 3(C1, C2), the marginalized posterior probability distributions and confidence bounds of the step height and width are presented for the simulated step heights of 9.92 nm, 19.84 nm, and 39.69 nm. The confidence bounds have been calculated as the 2.5% and 97.5% quantiles of the estimated distribution, which correspond to 2-sigma confidence bounds for the Gaussian distribution.

Calculated probability distributions resemble Gaussian distributions. Confidence bounds in all cases are tight and lie within the prior space bounds (see Table 1). The true values of parameters lie within confidence bounds, although in some cases, they are very close to their endpoints. Those observations indicate the correctness of the posterior probability distributions calculated by DNS.

From Eq. 7, the parameters with a high posterior probability provide a low mean squared residual of the modelled fringe pattern. To analyze this aspect, we calculated the mean values of the obtained posterior distributions for step heights of 9.92 nm, 19.84 nm, and 39.69 nm as the estimated values of the model parameters. For those values, we calculated the fitted fringe patterns and the residual maps. The respective results are presented in Figure 3(C3-C5). As the noise is a main factor contributing to the regression MSR, the residual maps represent the difference between the fitted and noise-free fringes.

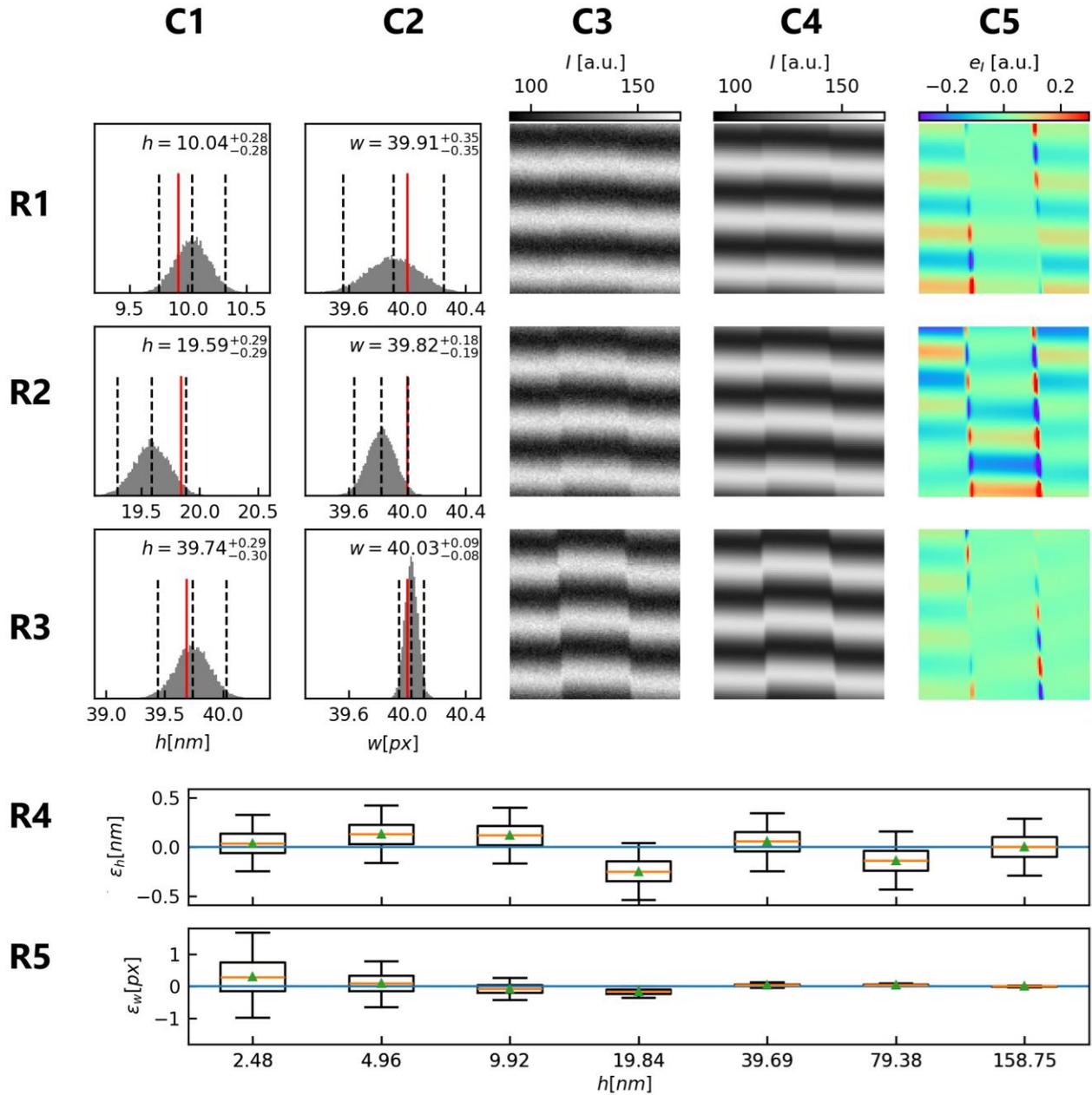

*Figure 3. Simulation results for the varying step heights. The detailed results for step heights 9.92 nm, 19.84 nm, and 39.69 nm are presented in rows R1, R2, and R3, respectively. The marginalized probability distributions of step height and width are presented in columns C1 and C2, respectively. Dashed lines represent mean values and confidence bounds. Red lines represent true values of parameters. In columns C3, C4, and C5, respectively, the simulated noisy fringe*

*pattern, the fitted fringe pattern and the noise-excluded residual map are presented. The box plots of marginalized probability distributions of estimation errors for step height and width are presented in rows R4 and R5, respectively. Boxes indicate 25% and 75% distribution quantiles. Whiskers represent 2-sigma quantiles. Orange lines and green triangles represent median and mean values of the error distributions.*

The numerical results confirm the link between high posterior probability and low regression MSR. The noise-excluded residuals across the image are much smaller than the fringe amplitude. The most significant residuals appear at the edges of the step structure, and this is a consequence of the discontinuous character of the step object.

In Figure 3 (R4, R5), the box plots of marginalized probability distributions of estimation errors for step height and width are presented. The results cover a wide range of step heights varying from 2.48 nm to 158.75 nm. We doubled the simulated step height for each subsequent sample. For the assumed light wavelength of 635 nm, the step heights introduce phase shifts $\varphi_h$ from $\pi/64$ rad to $\pi$ rad (see Eq. 10). Those phase shift in turn results in shifts between background and step fringe patterns varying from 0.25 px to 16 px. The shift between background and step fringe patterns in pixels can be calculated from

$$\Delta p = \frac{\varphi_h}{2\pi} \cdot T = \frac{2h}{\lambda} \cdot T \qquad (11)$$

It can be noticed that, for the step width, the spread of the error distribution decreases with the step height. That can be explained by the increased distinguishability of the step from the background for higher steps. Another behavior can be observed for the error distributions of the step height. The distribution spread is constant, and the mean values vary randomly across the tested range of the step heights. Random changes of mean distribution values result from the actual spatial distribution of the noise signal within the fringe pattern.

The presented results show the capability of our framework to measure the steps of heights ranging from single nanometers to more than a hundred nanometers. Even higher steps can be measured, but for $h \geq \frac{\lambda}{2}$ the non-injectivity of the cosine function in Eq. 4 must be considered.

We also investigated the challenging case of the subpixel shift of the step fringes. We ran the algorithm for the step heights 0.62 nm, 1.24 nm, and 2.48 nm, which correspond to the 0.0625 px, 0.125 px and 0.25 px shifts between step and background area (see Eq. 11). The results are presented in Supporting Information S3. DNS is capable of returning meaningful results also in that case, but the prior intervals must be chosen carefully to capture the whole domain of the posterior distribution. Also, the confidence intervals may become excessively wide for low step heights, indicating high uncertainty of the estimated parameters.

Next, we have evaluated the estimation algorithm for the varying Gaussian noise level. The step height was set to 19.84 nm, the fringe period was set to 32 px, and the noise standard deviation was changed from 3 to 30, which corresponds to decreasing the signal-to-noise ratio ($SNR = a/\sigma$) from 10 to 1 ($a = 30$). The box plots of the estimation errors for step height and width are presented in Figure 4.

For the increasing noise level, the spreads of the error distributions increase, which can be explained by the decreasing distinguishability of the step from the background. Similarly to the varying step height simulations, the mean values of the error distributions change randomly around zero, suggesting their dependence on the spatial distribution of the noise signal across the fringe pattern. In the tested cases, the computed probability distributions lie entirely within the prior bounds for $\sigma \leq 24$, which correspond to SNR as low as 1.25. For the discussion on DNS behavior for even lower SNR see Supporting Information S3.

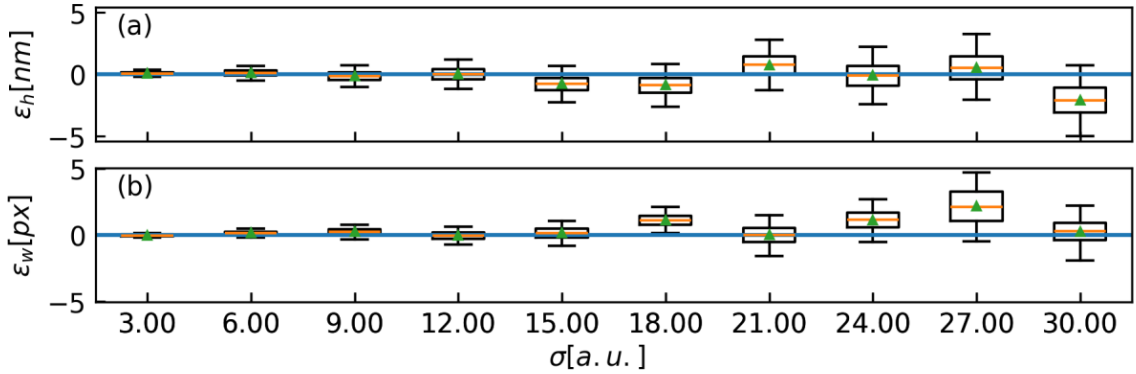

*Figure 4. Box plots of marginalized probability distributions of estimation errors against noise level for step height (a) and width (b). Boxes indicate 25% and 75% distribution quantiles. Whiskers represent 2-sigma quantiles. Orange lines and green triangles represent the median and mean value of the error, respectively.*

Finally, we have evaluated the estimation algorithm for the varying fringe period (Figure 5). The step height was set to 19.84 nm, and the noise standard deviation was set to 0.1. The fringe period was changed from 6 px to 384 px, which corresponds to decreasing the number of fringes from 16 fringes to a quarter of the fringe within the field of view.

The spread of error distributions remains constant for both step parameters and for fringe periods up to 192 px, with a small increase in the spread for the largest period 384 px. The results show that DNS performs equally well, regardless of the fringe period. The applicability of the method to low fringe frequencies is especially valuable, as some optical systems are limited to this range due to factors such as the use of partially coherent light sources (in common-path systems valued for their robustness and improved SNR) or the need to maximize the system's space-bandwidth product usage. For further discussion, see Supporting Information S3.

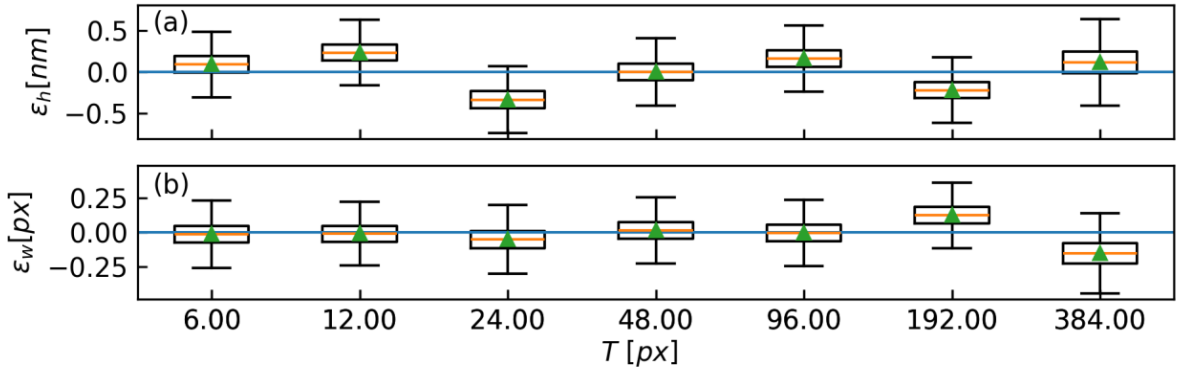

*Figure 5. Box plots of marginalized probability distributions of estimation errors against the fringe period for step height (a) and width (b). Boxes indicate 25% and 75% distribution quantiles. Whiskers represent 2-sigma quantiles. Orange lines and green triangles represent the median and mean value of the error, respectively.*

The DNS computations were performed on a virtual machine equipped with Ubuntu 24.04.2 LTS operating system (Oracle VirtualBox 7.0, Windows 11 host). We used an open-source Python implementation of the DNS algorithm provided by the *dynesty* library (version 2.1.5) [18]. The library supports multithreading, and four threads were used for computations. The utilized CPU was AMD Ryzen 7 3700X 3.6GHz. On average, DNS execution in the presented simulations took about 15 minutes.

EXPERIMENTAL RESULTS. We have evaluated the estimation algorithm for the calibration step object with the metrologically certified (by Polish Central Office of Measures) height $15.0^{+1.0}_{-1.0}$ nm. The fringe pattern image size is 500 × 500 px. The selected prior bounds, based on visual inspection of the fringe pattern, are presented in Table 2.

The calculated probability distributions of the step height and width for the calibration object are presented in Figure 6(a, b). The obtained distributions are Gaussian-like with tight confidence bounds lying entirely within the prior bounds. The estimated step height $14.886^{+0.089}_{-0.090}$ nm lies within the certified height $15.0^{+1.0}_{-1.0}$ nm of the tested object.

Table 2. Prior intervals of model parameters for the calibration object measurements.

| Parameter | $\varphi_h[\text{rad}]$ | $w[\text{px}]$ | $p_x[\text{px}]$ | $p_y[\text{px}]$ | $\beta[°]$ |
|---|---|---|---|---|---|
| **Prior interval** | $[0, 0.5\pi]$ | $330 \pm 5$ | $const = 250$ | $240 \pm 5$ | $90 \pm 5$ |
| **Parameter** | $\delta[°]$ | $\varphi_l[\text{rad}]$ | $T[\text{px}]$ | $a[a.u.]$ | $b[a.u.]$ |
| **Prior interval** | $0.0 \pm 2.5$ | $[0, \pi]$ | $45.5 \pm 2.5$ | $30 \pm 5$ | $130 \pm 5$ |

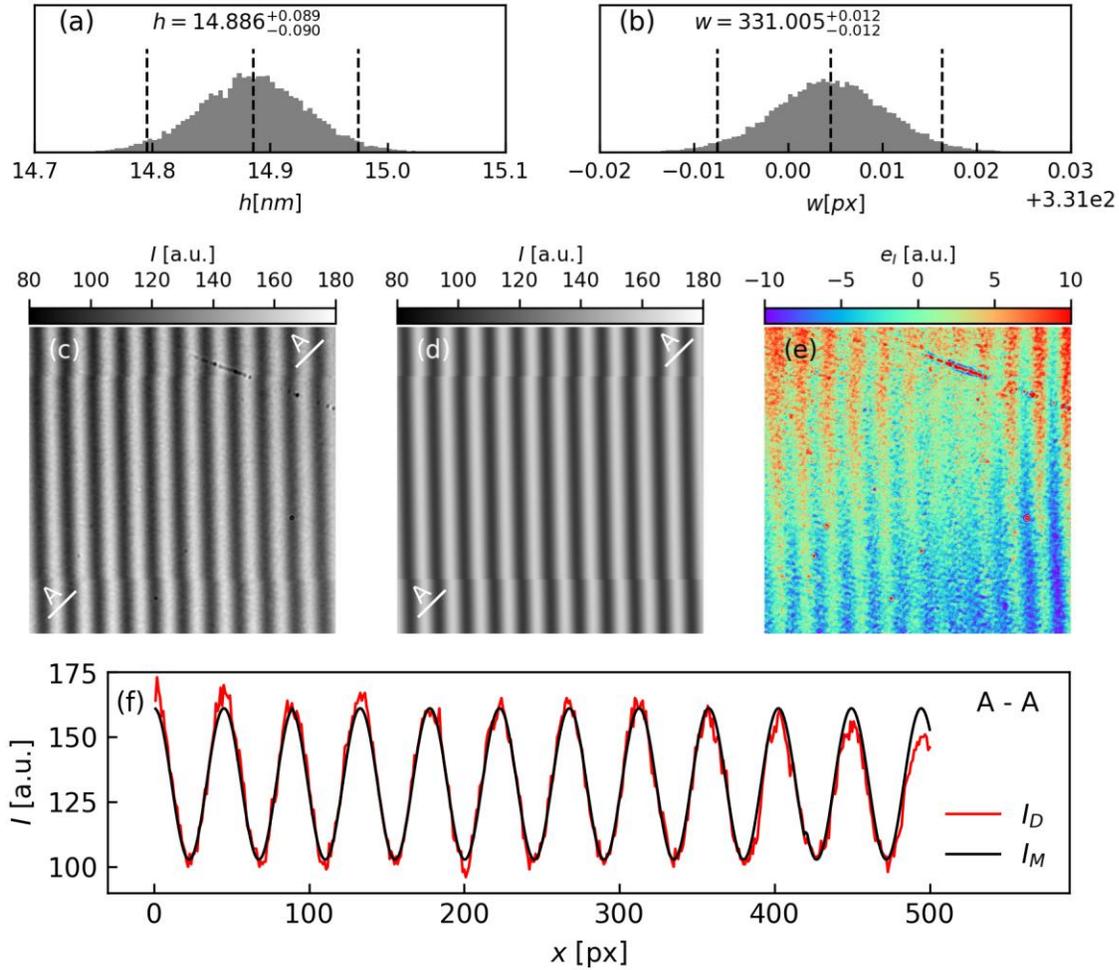

*Figure 6. Experimental results for the calibration step object. Probability distributions of the estimated step height (a) and width (b). Measured fringe pattern (c), fitted fringe pattern (d), regression residual map (e). Comparison of diagonal cross section A-A through the measured and fitted fringe patterns (f).*

In Figure 6(c, d, e) the measured fringe pattern is compared with the modelled fringe pattern, calculated for the mean-value estimates of model parameters derived from the posterior probability distribution. Although the fringe pattern is visually fitted correctly, a sinusoidal pattern is clearly visible in the residual map. In Figure 6(f), the diagonal cross sections of the measured and fitted fringe patterns are compared. The sinusoidal residual pattern results from the variations of the fringe amplitude and mean intensity across the image. Nevertheless, the fringe period, initial phase shift and orientation are estimated correctly. The regression results show the robustness of the estimation algorithm to the imperfections in the registered data, such as noise, nonuniform intensity across the image, or small defects resulting from dust particles residing on optical surfaces of the experimental setup.

For the image of size 500×500, the execution of DNS lasted 9.6 hours. To assess the quality of parameter estimation with a reduced amount of data, we downsampled the fringe pattern to a size of 100×100 pixels and repeated the estimation procedure. Estimation results for the downsampled data are presented in Supporting information S4. For the downsampled fringe pattern, DNS still returns reliable results (step height $14.964^{+0.682}_{-0.699}$ nm lying within the certified object height $15.0^{+1.0}_{-1.0}$ nm), with the execution time significantly reduced to 13 minutes.

As the second experimental case, we estimated the parameters of the waveguide object with the design height equal to 8 nm. The optical waveguide was made of silicon nitride material. The silicon nitride based optical waveguide has gained significant interest as it supports transmission of the visible light, has high refractive index contrast and it is compatible with standard CMOS fabrication. A shallow rib waveguide geometry was used, featuring a rib height of only 8 nm and a slab height of approximately 150 nm. The details of design and fabrication of single mode rib waveguide is discussed in previous literature ([31, 32]). Shallow rib waveguides, with rib heights of 2–8 nm as used in this work, are essential to design single mode condition for high-refractive index contrast optical waveguide (Silicon nitride) in the visible spectrum.

The measured fringe image of size 281 × 961 px is presented in Figure 7(c) (waveguide area is marked with an arrow). A mask was used to extract the region of interest. The selected prior bounds, based on visual inspection of the fringe pattern, are presented in Table 3.

Table 3. Prior intervals of model parameters for the waveguide object.

| Parameter | $\varphi_h[rad]$ | $w[px]$ | $p_x[px]$ | $p_y[px]$ | $\beta[°]$ |
|---|---|---|---|---|---|
| Prior interval | $[-\frac{\pi}{4},\frac{\pi}{4}]$ | $20 \pm 5$ | $const = 430.5$ | $145 \pm 5$ | $95 \pm 5$ |
| Parameter | $\delta[°]$ | $\varphi_l[rad]$ | $T[px]$ | $a[a.u.]$ | $b[a.u.]$ |
| Prior interval | $0.0 \pm 5$ | $[-\frac{3\pi}{2},-\frac{\pi}{2}]$ | $330 \pm 30$ | $825 \pm 25$ | $1850 \pm 50$ |

The waveguide is made of a transparent dielectric material (silicon nitride); thus, during measurement, the object beam entered the waveguide and was reflected from its base. In such case, the step height $h$ can be calculated from the step induced phase shift $\varphi_h$ using the formula:

$$h = \frac{\lambda \cdot \varphi_h}{4\pi \cdot (n-1) \cdot \cos\theta}, \tag{12}$$

where $\lambda = 620\ nm$, refractive index of the silicon nitride waveguide material $n = 2.0411$ and $\theta = \text{asin}\ NA$ for $NA = 0.55$.

The calculated probability distributions of the step height and width for the waveguide are presented in Figure 7(a, b). In Figure 7(c, d, e), the measured fringe pattern is compared with the modelled fringe pattern, calculated for the mean-value estimates of model parameters.

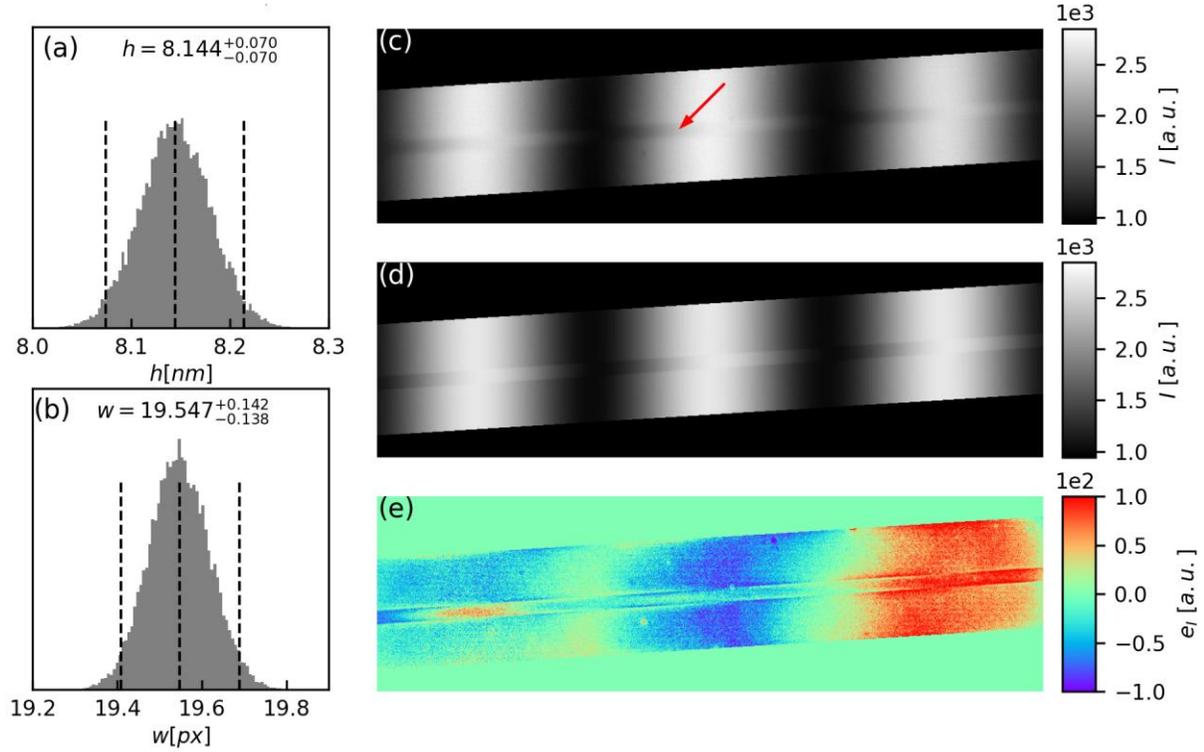

*Figure 7. Experimental results for the waveguide object. Probability distributions of the estimated step height (a) and width (b). Measured fringe pattern (c), fitted fringe pattern (d), and regression residual map (e).*

Similarly to the calibration object measurement, we notice a sinusoidal pattern in the residual map, which is a result of the nonuniform mean intensity level and fringe amplitude across the image. Despite that, the waveguide placement defined by its width, position and orientation, is estimated correctly. However, the initial DNS run returned the step height of 7.326 nm, which is significantly underestimated compared to the design height 8 nm. We have determined that the source of that inconsistency is a relatively wide transition area (from the top of the waveguide to the ground) compared to the width of the waveguide. To obtain a reliable estimate of the waveguide top height, we performed a second, fine-tuning DNS run, with a masked transition region. Masking is realized by setting the residuals in the transition region to zero. The refined step height is $8.144^{+0.07}_{-0.07}$ nm, which is much closer to the design height 8 nm.

To confirm the correctness of the refined results, we reconstructed the phase map using the Temporal Phase Shifting (TPS) algorithm followed by phase unwrapping. The waveguide top height calculated from the TPS-reconstructed phase map is 8.178 nm and it lies within the refined confidence bounds $8.144^{+0.07}_{-0.07}$ nm. For the discussion of the transition region residuals, the fine-tuning procedure, and TPS results, see Supporting Information S5.

While the basic DNS run took 9.1 hours, the fine-tuning DNS run required only 1.2 hours due to the reduced dimensionality of the prior space.

**DISCUSSION**

The simulation and experimental results confirm the notable advantages of the presented Bayesian framework for single-shot widefield measurements of nano-scale elements. The proposed method achieves high accuracy and sensitivity across a broad range of simulated measurement conditions, also in the scenarios challenging for alternative methods, e.g. low distinguishability of the object, low SNR, and uniform field mode. The experimental results highlight the robustness of the proposed method to the imperfections of the measurement data – the feature which gives a potential to reduce requirements (and cost) for the measurement system. The employed probabilistic framework provides complete measurement results, including uncertainty assessment, thus eliminating the need for additional statistical analysis.

While the framework offers remarkable advantages, the main issue is the relatively long computation time required to estimate the posterior probability distribution over the entire prior domain. The region of high posterior mass typically represents only a small subset of the prior domain, which leads to high computational costs when trying to localize it. To address this issue, future research will explore procedures to precisely localize the typical set before the main parameter estimation is performed, streamlining the process.

Another area to consider is the need for a representative model of the measured object. For example, assuming a rectangular shape for the waveguide resulted in an underestimation of the step height parameter. However, the flexibility of the framework is a significant advantage here – one can easily compare and validate a range of models tailored to the specific needs. Modeling other basic geometric shapes is straightforward, and for more complex or unknown shapes, advanced techniques such as splines or Zernike polynomials can be employed to ensure precision.

This study establishes Bayesian inference as a powerful tool for semiconductor nanometrology, bridging the gap between theoretical statistical modeling and practical widefield optical measurement techniques. As the demand for precision in semiconductor manufacturing continues to rise, probabilistic metrology frameworks like the one proposed here are poised to play an increasingly vital role in ensuring the reliability and scalability of nanoscale device production and characterization. For example, optical waveguides with 2 nm rib heights, e.g. in [33], have been used for different applications. As surface roughness for such waveguides is typically below 1 nm, stringent requirements on measurement precision are imposed, to be secured by the proposed method. Our findings underscore the transformative potential of Bayesian inference in the field of non-scanning widefield optical nanometrology, setting the stage for further exploration and application in diverse areas of semiconductor technology, from photonic integrated circuits to advanced quantum devices.

**METHODS**

For measurements described in Figure 6 we developed a Linnik interferometer, Figure 8, with accurate vertical scanning in the object arm and the piezoelectric transducer (PZT) controlling the position of the reference mirror. The measurement method combines the use of an interferometric system, and numerical techniques of analysis and image processing. The low-coherence light source used is an LED ($\lambda$ = 635 nm, FWHM = 15 nm). The beam emerging from the end of a multimode fiber - acting as a quasi-point light source (PS) - is collimated by collimator C. It is then shaped by lens L1 and split by a beam splitter cube (BS) into the object and reference beams.

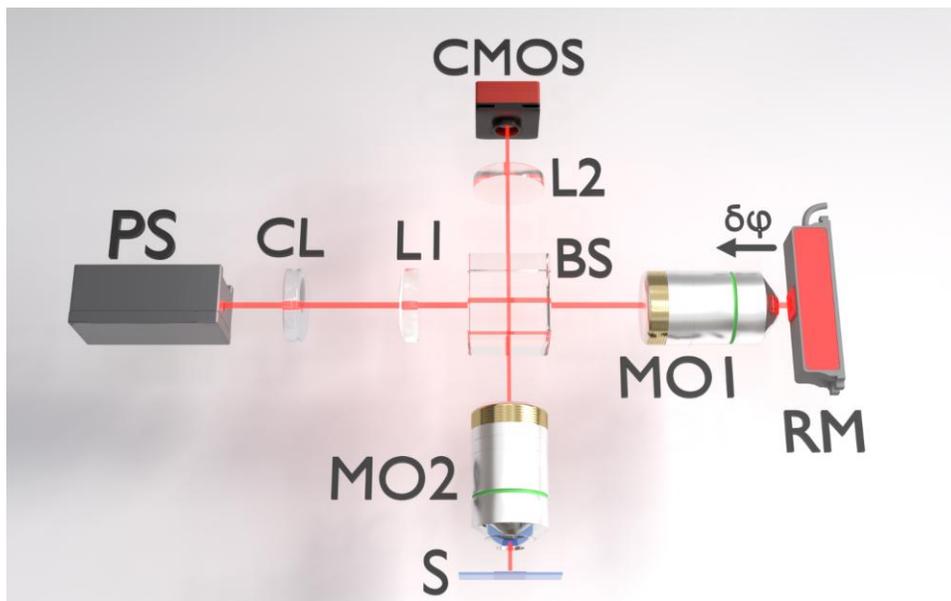

*Figure 8. Scheme of the Linnik interferometer used in experimental works.*

The reference beam passes through an afocal system consisting of lens L1 (focal length 200 mm) and microscope objective MO1, and is directed onto a flat reference mirror (RM). The object beam is formed by a second afocal system (L1 and microscope objective MO2) and illuminates the sample under test. Both microscope objectives are Nikon PLAN APO 40x/0.95.

The reflected beams return through their respective afocal imaging systems: MO1 and lens L2 for the object arm, and MO2 and lens L2 for the reference arm. The resulting interference pattern is captured by a CMOS camera (BASLER acA1920-150um, resolution 1920 × 1200, pixel size 4.8 μm).

Using a low-coherence LED light source and a low-noise camera is a fundamental physical approach to minimizing noise in the interferometric system. In our setup, the standard deviation of noise is approximately 1 nm. Due to the limited temporal and spatial coherence of the LED, high-quality fringes - with contrast up to 50% of the maximum - are obtained at a distance of 9 μm.

The interferogram for the optical rib waveguide object, analyzed in Figure 7, has been recorded using an optical configuration of the Linnik interference microscopy system with pseudothermal light source [34, 35].

ASSOCIATED CONTENT

**Supporting Information**. The supporting information is available. (S1) Explanation of the ambiguity of the step position choice in the distance function definition. (S2) Derivation of the likelihood term for the regression task. (S3) Detailed simulation results. (S4) Parameters estimation for the calibration object and downsampled fringe pattern. (S5) Analysis of the transition region residuals for the waveguide parameters estimation.

AUTHOR INFORMATION

**Corresponding Authors**

Damian M. Suski – Warsaw University of Technology, Institute of Automatic Control and Robotics, 8 Sw. A. Boboli St., 02-525, Warsaw, Poland, ORCID orcid.org/0000-0001-6389-5428, e-mail: damian.suski@pw.edu.pl

Maria Cywińska – Warsaw University of Technology, Institute of Micromechanics and Photonics, 8 Sw. A. Boboli St., 02-525, Warsaw, Poland, ORCID orcid.org/0000-0001-6246-2727, e-mail: maria.cywinska@pw.edu.pl

Maciej Trusiak – Warsaw University of Technology, Institute of Micromechanics and Photonics, 8 Sw. A. Boboli St., 02-525, Warsaw, Poland, ORCID orcid.org/0000-0002-5907-0105, e-mail: maciej.trusiak@pw.edu.pl

**Author Contributions**

‡ Damian Suski and Maria Cywińska contributed equally.

**Notes**

The authors declare no conflict of interest.

**ACKNOWLEDGMENTS**

This research was funded by the National Science Center Poland (MT: PRELUDIUM BIS 2022/47/O/ST7/02360, MC: OPUS 2024/55/B/ST7/02085) and Warsaw University of Technology within the Excellence Initiative: Research University (IDUB) programme (YOUNG PW Grant no. 504/04496/1143/45.010012). Balpreet Singh Ahluwalia acknowledge the support of Norwegian Directorate for Higher Education and Skills (Diku) project number UTF-2024/10439 and INCP2-2024/10233.

# Supporting Information for: Bayesian inference for precise single-shot widefield interferometric geometrical nanometrology

[1,4,‡]*Damian M. Suski,* [2,5,‡]*Maria Cywińska,* [2]*Julianna Winnik,* [2]*Michał Józwik,* [2]*Piotr Zdańkowski,* [3]*Azeem Ahmad,* [3]*Balpreet S. Ahluwalia,* [2,*]*Maciej Trusiak*

[1] Warsaw University of Technology, Institute of Automatic Control and Robotics,

8 Sw. A. Boboli St., 02-525, Warsaw, Poland

[2] Warsaw University of Technology, Institute of Micromechanics and Photonics,

8 Sw. A. Boboli St., 02-525, Warsaw, Poland

[3] UiT The Arctic University of Norway, Department of Physics and Technology,

35 Klokkargårdsbakken St., 9019, Tromsø, Norway

[4] damian.suski@pw.edu.pl

[5] maria.cywinska@pw.edu.pl

* maciej.trusiak@pw.edu.pl

‡authors contributed equally

## S1. EXPLANATION OF THE AMBIGUITY OF THE STEP POSITION CHOICE IN THE DISTANCE FUNCTION DEFINITION

The definition of the distance function is:

$$d(x, y; \Theta_s) = -\cos\beta \cdot (x - p_x) + \sin\beta \cdot (y - p_y). \tag{S1.1}$$

The value of a distance function for any point $(x_c, y_c)$ lying on the center line of the step is equal to zero:

$$d(x_c, y_c; \Theta_s) = -\cos\beta \cdot (x_c - p_x) + \sin\beta \cdot (y_c - p_y) = 0. \tag{S1.2}$$

It means that the distance function will return exactly the same results if we replace the point $(p_x, p_y)$ with any point $(x_c, y_c)$ lying on the center line, because:

$$-\cos\beta \cdot (x - x_c) + \sin\beta \cdot (y - y_c)$$

$$= -\cos\beta \cdot (x - x_c + p_x - p_x) + \sin\beta \cdot (y - y_c + p_y - p_y)$$

$$= -\cos\beta \cdot (x - p_x) + \sin\beta \cdot (y - p_y) - \underbrace{[-\cos\beta \cdot (x_c - p_x) + \sin\beta \cdot (y_c - p_y)]}_{=0}$$

$$= -\cos\beta \cdot (x - p_x) + \sin\beta \cdot (y - p_y). \tag{S1.3}$$

## S2. DERIVATION OF THE LIKELIHOOD TERM FOR THE REGRESSION TASK

In the regression task, the standard assumption is that the regression residual at every measurement point is normally distributed with a zero mean and the standard deviation $\sigma$:

$$e(x, y; \Theta) = I_M(x, y; \Theta) - I_D(x, y) \sim \mathcal{N}(0, \sigma^2). \tag{S2.1}$$

It means that at each pixel the probability density of the residual is:

$$p(e(x, y)|\Theta) = \frac{1}{\sqrt{2\pi\sigma^2}} \exp\left(-\frac{e^2(x, y; \Theta)}{2\sigma^2}\right). \tag{S2.2}$$

Assuming independence of the residuals at different pixels, the probability density of the residual distribution across all pixels (so the sought likelihood) is the product of probability densities at each pixel:

$$\mathcal{L}(\Theta) = p(e|\Theta) = \left(\frac{1}{\sqrt{2\pi\sigma^2}}\right)^N \cdot \exp\left(-\frac{\Sigma_{x,y}\, e^2(x,y;\Theta)}{2\sigma^2}\right), \quad (S2.3)$$

where $N = s_x \cdot s_y$ is a number of pixels in the image of size $s_x \times s_y$. If we calculate the partial derivative of $\mathcal{L}(\Theta)$ with respect to $\sigma$ we get:

$$\frac{\partial}{\partial \sigma}\mathcal{L}(\Theta) = \left(-\frac{N}{\sigma} + \frac{\Sigma_{x,y}\, e^2(x,y;\Theta)}{\sigma^3}\right) \cdot \left(\frac{1}{\sqrt{2\pi\sigma^2}}\right)^N \cdot \exp\left(-\frac{\Sigma_{x,y}\, e^2(x,y;\Theta)}{2\sigma^2}\right)$$

$$= \left(-\frac{N}{\sigma} + \frac{\Sigma_{x,y}\, e^2(x,y;\Theta)}{\sigma^3}\right) \cdot \mathcal{L}(\Theta). \quad (S2.4)$$

By comparing the above derivative with zero and noting that $\mathcal{L}(\Theta) > 0$, we see that the likelihood has its maximum with respect to $\sigma$ for $\sigma^2$ equal to the mean squared residual (MSR):

$$\sigma^2 = \frac{1}{N}\Sigma_{x,y}\, e^2(x,y;\Theta). \quad (S2.5)$$

In such case, the likelihood formula takes the form:

$$\mathcal{L}(\Theta) = (2\pi\sigma^2)^{-\frac{N}{2}} \cdot \exp\left(-\frac{N}{2}\right)$$

$$= \exp\left(-\frac{N}{2} \cdot (\log(2\pi\sigma^2) + 1)\right). \quad (S2.6)$$

### S3. DETAILED SIMULATION RESULTS

SUBPIXEL SHIFT CASE. To investigate the estimation results for the challenging subpixel shift of step fringes, we ran the algorithm for the step heights 0.62 nm, 1.24 nm, and 2.48 nm, which correspond to the 0.0625, 0.125, and 0.25 pixel shifts between step and background area. In Figure S3.1(C1, C2), the marginalized probability distributions of step height and width are presented.

For the 0.0625px shift, the probability distribution for the step width does not fit within the assumed prior interval of $40 \pm 5$px. For the 0.125 px shift, the width distribution still does not lie entirely within the prior bound, but it becomes Gaussian-like with a clearly recognizable peak. For the 0.25 px shift, the width distribution is Gaussian-like and it lies entirely within the assumed prior interval. Interestingly, in all cases, the distribution of step height is Gaussian-like, and the true height lies within the assumed confidence bounds.

The mean value of the posterior distribution, which does not lie within the assumed prior interval, cannot be treated as reliable estimate of the model parameter. Nevertheless, we may notice that in all cases the true width value lies in the so-called typical set [S1], i.e. the area capturing most of the distribution's posterior mass.

The simulated fringe patterns are presented in Figure S3.1(C3). For the considered step heights, the step area is not distinguishable from the background area. For the mean-value estimates of model parameters, the modelled fringe pattern as well as the residual maps were calculated (Figure S3.1(C4,C5)). Although the residuals are small, the step width estimates are subject to high uncertainties due to the low distinguishability of the step from the background.

From the above observations, we claim that DNS is capable of returning meaningful results also for the subpixel cases, but the prior intervals must be chosen carefully to capture whole domain of the posterior distribution. Also, the lengths of confidence intervals increases with the decreasing distinguishability of the step from the background and may become excessively large for very low step heights.

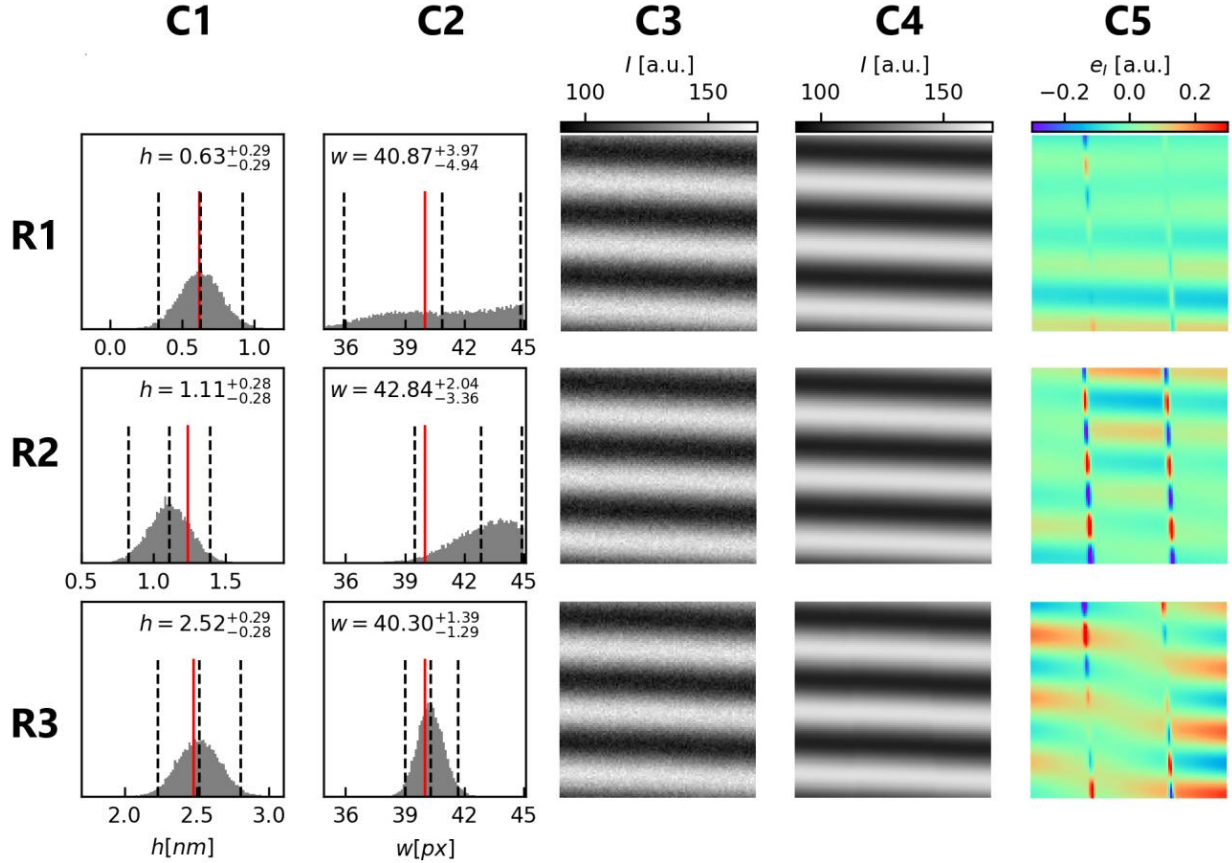

*Figure S3.1. Simulations results for the step heights 0.62 nm (R1), 1.24 nm (R2), and 2.48 nm (R3), corresponding to subpixel shifts between step and background fringes. The marginalized probability distributions of step height and width are presented in columns C1 and C2. Dashed lines represent mean values and confidence bounds. Red lines represent true values of parameters. In columns C3, C4, and C5, the simulated noisy fringe pattern, the fitted fringe pattern and the noise-excluded residual map are presented.*

VARYING NOISE LEVEL CASE. In Figure S3.2 we present the detailed simulation results for noise standard deviations $\sigma = 15; 21; 27$ ($h = 19.84\ nm, T = 32\ px, a = 30$), which correspond to $SNR = 2;\ 1.43;\ 1.11$.

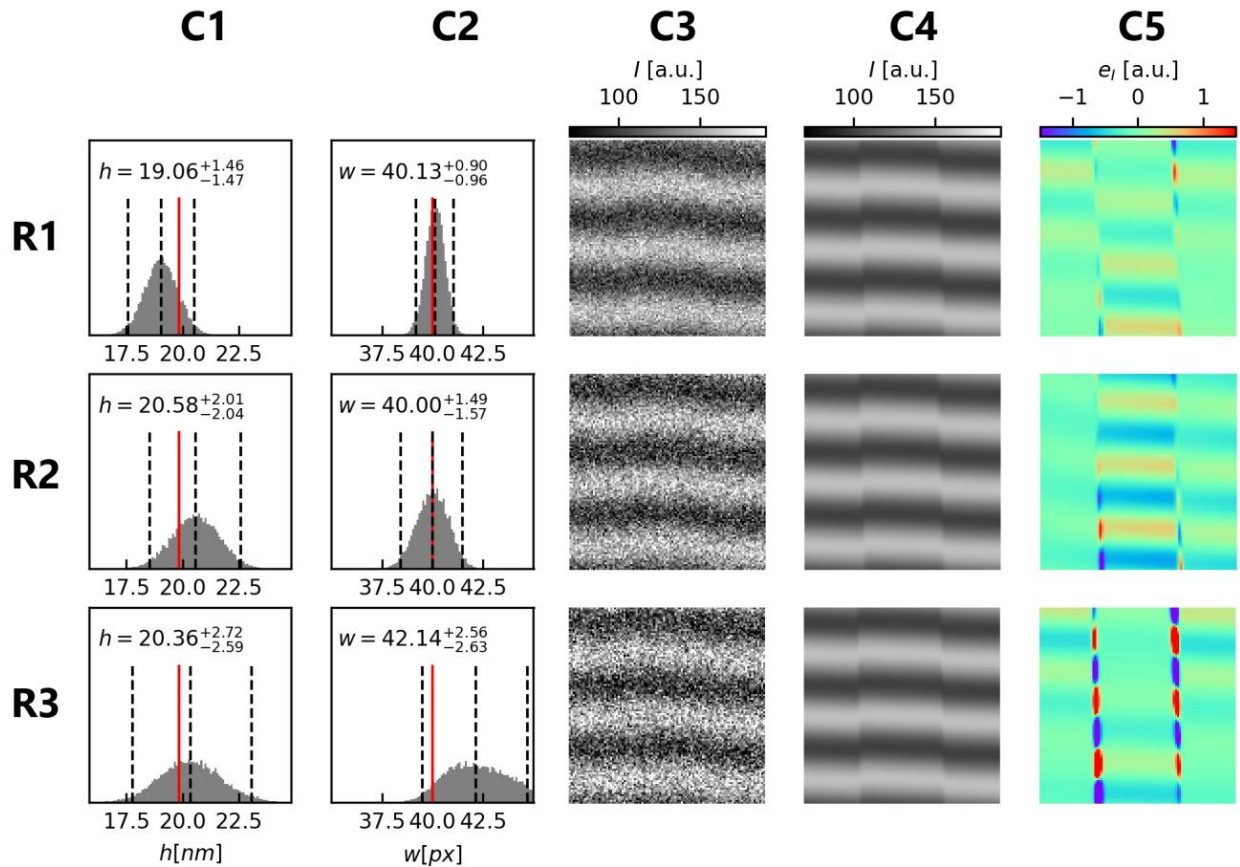

*Figure S3.2. Simulations results for the noise standard deviation 15 (R1), 21 (R2), and 27 (R3), The marginalized probability distributions of step height and width are presented in columns C1 and C2. Dashed lines represent mean values and confidence bounds. Red lines represent true values of parameters. In columns C3, C4, and C5, the simulated noisy fringe pattern, the fitted fringe pattern and the noise-excluded residual map are presented.*

We notice that the width distribution for $\sigma = 27$ does not fit within the assumed prior interval $40 \pm 5 \, px$. Nevertheless, DNS correctly captures the distribution shape limited to the prior bounds. Similarly to the subpixel shift case, if the prior bounds cover the whole typical set of the posterior distribution, DNS is capable of returning reliable estimation results.

Even though, at higher noise levels, the step area is hardly distinguishable from the background (Figure S3.2(C3)), the mean value estimates of model parameters still provide a good fit between the modeled and simulated fringe pattern (Figure S3.2(C4, C5)).

VARYING FRINGE PERIOD. In Figure S3.3 we present the detailed simulation results for fringe periods $T = 6; 48; 384\ px$ ($h = 19.84\ nm, \sigma = 0.1$).

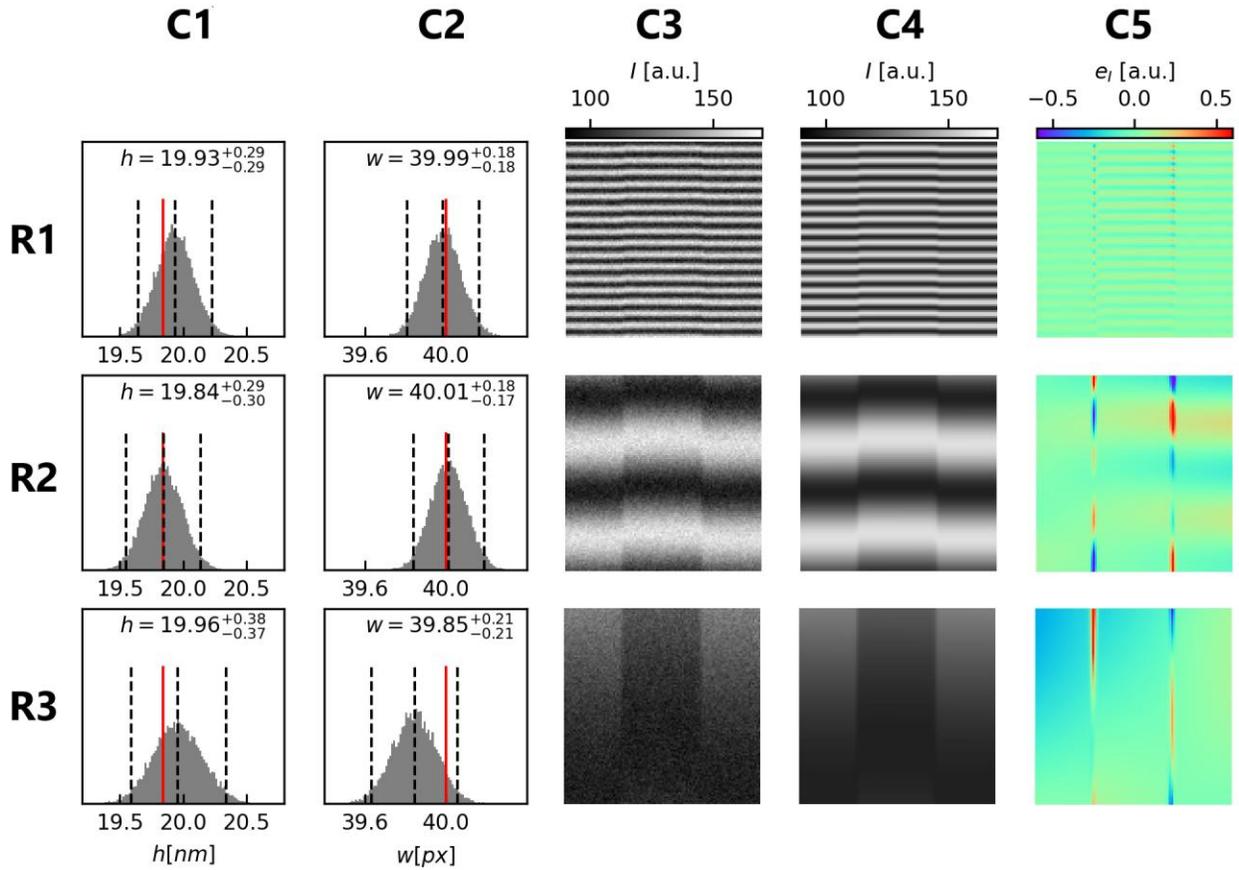

*Figure S3.3. Simulations results for the fringe period 6 px (R1), 48 px (R2), and 384px (R3). The marginalized probability distributions of step height and width are presented in columns C1 and C2. Dashed lines represent mean values and confidence bounds. Red lines represent true values of parameters. In columns C3, C4, and C5, the simulated noisy fringe pattern, the fitted fringe pattern and the noise-excluded residual map are presented.*

The distributions' spreads are the same for $T = 4$ px and $T = 48$ px. For $T = 384$ px, the height distribution spread increases by 30% and the width distribution spread by 24%. In all cases, the confidence bounds are tight and probability distributions lie entirely within the prior bounds.

The increase of the confidence bounds for $T = 384$ px is caused by the decreased distinguishability of the step area from the background in the lower part of the image (Figure S3.3(C3)). Nevertheless, in all cases, the mean value estimates of model parameters provide a good fit between the modeled and simulated fringe pattern (Figure S3.3(C4, C5)).

## S4. PARAMETERS ESTIMATION FOR THE CALIBRATION OBJECT AND DOWNSAMPLED FRINGE PATTERN

For the calibration object fringe pattern of size 500 x 500 pixels, the execution of DNS lasted 9.6 hours. To assess the quality of parameter estimation for the reduced amount of data, we have downsampled the fringe pattern to a size 100 x 100 pixels and repeated the estimation procedure. For the downsampled fringe pattern, the execution of the algorithm took 13 minutes.

The probability distributions of step height and width for the downsampled image case are presented in Figure S4.1(a, b). The obtained distributions are unimodal, but the skewness in the distributions of step width is apparent. The confidence bounds for all parameters increased significantly in comparison to original data. The estimated step height is $14.964^{+0.682}_{-0.699}$ nm in the downsampled image case, but it still lies within the certified height $15.0^{+1.0}_{-1.0}$ nm of the object.

In Figure S4.1(c, d, e) the downsampled fringe pattern is compared with the modelled fringe pattern, calculated for the mean-value estimates of model parameters. The residual map for the downsampled data case is similar to the residual map for the original data.

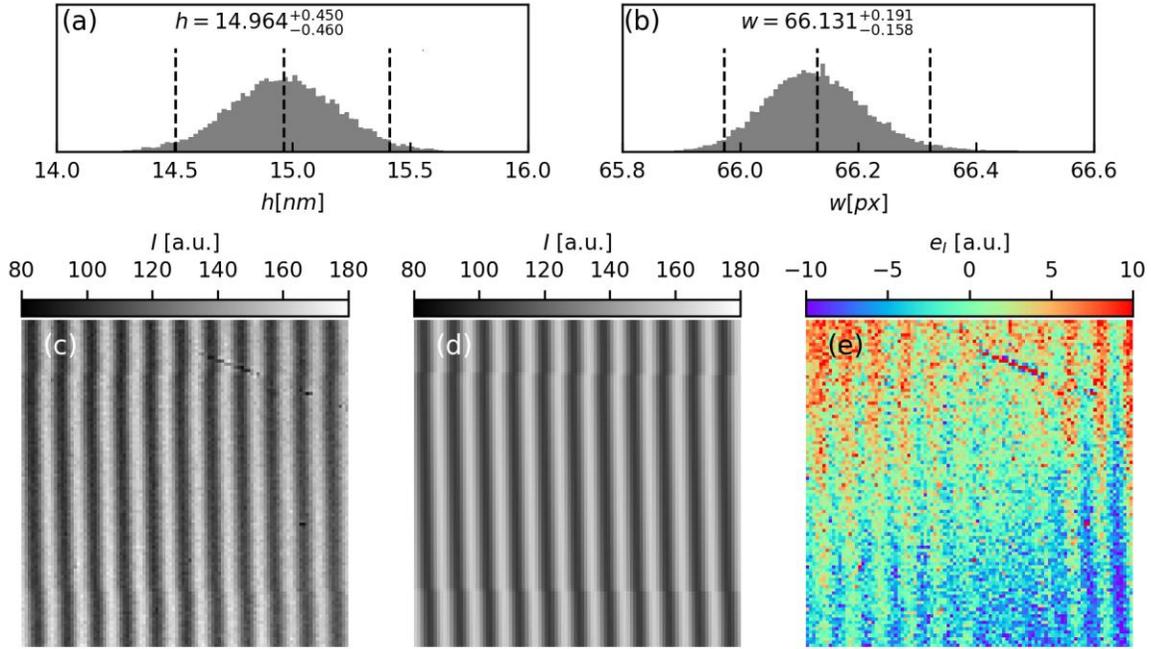

*Figure S4.1. Experimental results for the calibration step object and downsampled fringe pattern. Probability distributions of the estimated step height(a) and width(b). Measured fringe pattern (c), fitted fringe pattern (d), regression residual map (e).*

## S5. ANALYSIS OF THE TRANSITION REGION RESIDUALS FOR THE WAVEGUIDE PARAMETERS ESTIMATION

For the waveguide object, the initial DNS run returned the step height 7.326 nm, which is significantly underestimated in comparison to the design height 8 nm. To investigate the source of that inconsistency, we performed the phase reconstruction using the Temporal Phase Shifting (TPS) algorithm with phase unwrapping. The phase was reconstructed using five shifted fringe patterns, and the reconstructed phase map is presented in Figure S5.1(a). In Figure S5.1(b) the middle cross-section of the phase map is presented.

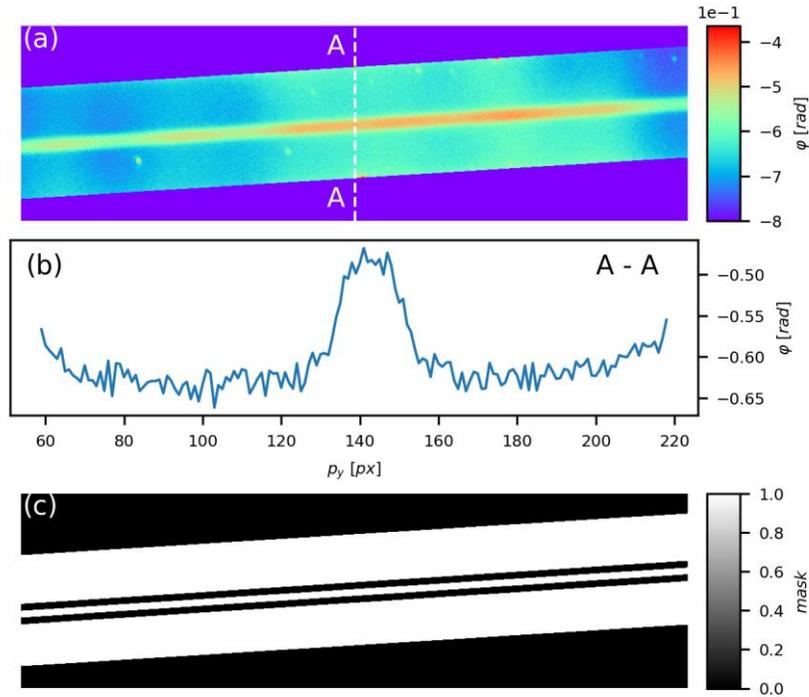

*Figure S5.1. Temporal phase shifting interferometry results for the waveguide object. The reconstructed phase map (a), cross-section A-A of the phase map (b), transition region mask (c).*

It can be noticed in the cross-section plot that the width of the transition region from the waveguide top to the ground is comparable to the width of the waveguide itself. That poses a question of how to correctly calculate the waveguide height from the phase map. On the basis of DNS estimation results, we created a mask presented in Figure S5.1(c), which covers the transition region. The step induced phase shift, was computed as the difference of the mean phase shifts calculated for the narrow waveguide top region and the ground region. The step height calculated from the TPS-reconstructed phase map with the masked transition region is 8.178 nm, which is close to the waveguide design height 8 nm.

The large slope of the waveguide sides, resulting in a wide region of transition from the ground to the waveguide top, was also expected to be a reason for the underestimation of the waveguide height by DNS. To confirm that, we ran again the estimation algorithm, but to eliminate the influence of the transition region residuals we masked the transition region, using the same mask as for TPS calculations. In the second algorithm run, we estimated only the phase shifts $\varphi_h$ and $\varphi_l$. The remaining parameters were kept constant at their mean value estimated from the first DNS run. The fine-tuned step height is $8.144^{+0.07}_{-0.07}$ nm, which agrees with TPS result (8.178 nm) and the designed waveguide height.

To investigate further the influence of transition region residuals on the waveguide parameter estimation we calculated the squared residuals maps for the basic and fine-tuning results. The respective residual maps are presented in Figure S5.2(a, b).

Next, we have calculated the map of differences between the squared residuals for basic and fine-tuned results $\Delta e_I^2$ (see Figure S5.2(c)). We may notice that along the vertical cross sections the squared residuals differences for the waveguide top are smaller than the differences for the transition region. To see it more clearly, we equalized the mean differences along vertical cross sections and presented them in Figure S5.2(d). During the basic estimation run, the algorithm prefers to reduce the residuals in the relatively wide transition region, at the cost of increased residuals in the waveguide top region. That leads to the underestimation of the step height for the narrow waveguide. During the fine-tuning run of the estimation algorithm, the transition region is masked and the algorithm finds an unbiased step height, which minimizes the residuals sum in the waveguide top region.

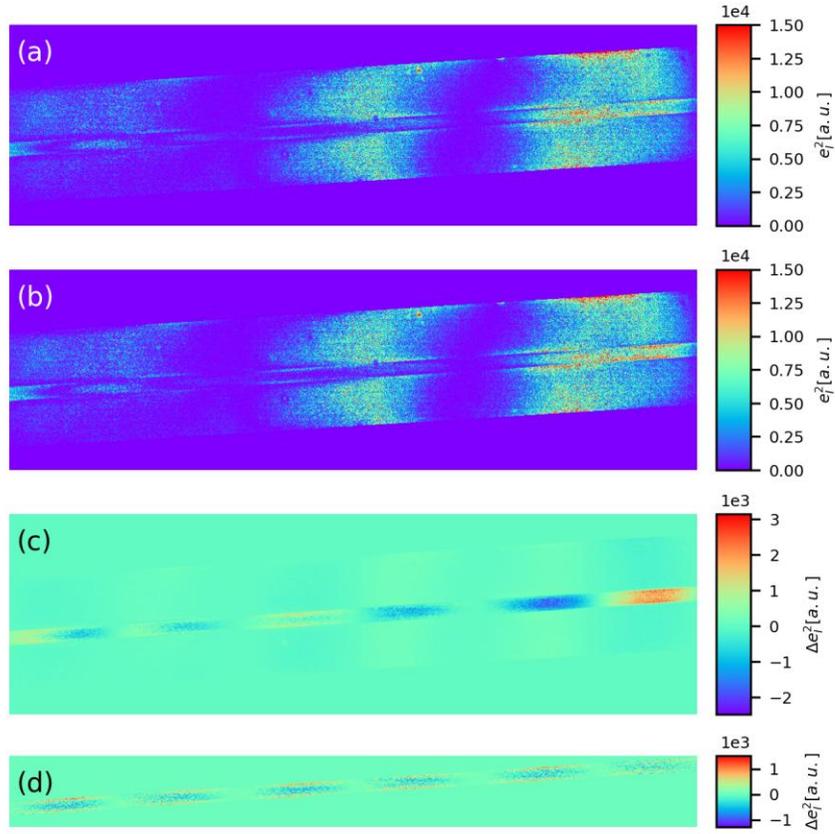

*Figure S5.2. Analysis of the regression residuals in the transition region. The squared residuals maps for the basic (a), and fine-tuned results (b). Map of the squared residuals differences between both runs - original (c), with equalized differences' means along vertical cross sections (d).*

**REFERENCES**

[S1] Joshua S. Speagle, "DYNESTY: a dynamic nested sampling package for estimating Bayesian posteriors and evidences", Mon. Not. R. Astron. Soc. 493, 3132-3158 (2020)